\documentclass[aps,pre,twocolumn,superscriptaddress]{revtex4-1}
\usepackage{graphicx}
\usepackage{bm}
\usepackage{epstopdf}
\usepackage{amsmath}
\usepackage{amsfonts}
\usepackage[usenames,dvipsnames]{color}
\usepackage{changes}
\usepackage[colorlinks=true,linkcolor=blue,urlcolor=blue,citecolor=blue]{hyperref}

\begin{document}
\title{Time scales in the thermal dynamics of magnetic dipolar clusters.}
\author{Paula Mellado}
\affiliation{School of Engineering and Sciences, 
	Universidad Adolfo Ib{\'a}{\~n}ez,
	Santiago, Chile
}

\begin{abstract}

The collective behavior of thermally active structures offers clues on the emergent degrees of freedom and the physical mechanisms that determine the low energy state of a variety of systems. Here, the thermally active dynamics of magnetic dipoles at square plaquettes is modeled in terms of Brownian oscillators in contact with a heat bath. Solution of the Langevin equation for a set of interacting $x-y$ dipoles allows the identification of the time scales and correlation length that reveal how interactions, temperature, damping and inertia may determine the frequency modes of edge and bulk magnetic mesospins in artificial dipolar systems.

\end{abstract}

\maketitle

\section{Introduction}
\label{sec:Introduction}
In the study of dynamical systems, temperature has long been an ally for the elucidation of new orders and phases of matter \cite{tauber2017phase,canfield2010feas,hermanns2018physics,hallas2018experimental}.  Yet, to capture the thermally active phenomenology of a system, a sense of timing is crucial. Therefore the accordance of the frequencies used in experimental probes, with a system proper time scales remains of particular interest   \cite{PhysRevLett.58.385}. Challenges arise because often there are several time scales, and worse still, one or few of them result from intrinsic interactions in the system \cite{topping2018ac,PhysRevE.93.032129,PhysRevX.1.021013}. A remarkable example is the case of cuprate metals where two transport relaxation times in the transport coefficients has been understood in terms of scattering processes that discriminate between currents that are even, or odd under charge conjugation \cite{coleman1996should}.

Taking full benefit of the experimental probes requires untangling of the dynamical response by  establishing a hierarchy of the proper time scales and associating to each of them concrete aspects of the system under analysis.  This, among other effects, facilitates the identification of tunable key parameters to guarantee that a complete thermal equilibrium of the system can be reached during the observation time.

The study of the dynamical relaxation and the response of a physical system to external fields is ubiquitous \cite{lu2013colloidal,banerjee2020actin,libchaber2019biology,henley2010coulomb}. Because of their distinctive behavior, here we choose to highlight the dynamics of frustrated  magnetic systems \cite{diep1994magnetic,PhysRevX.7.021030}. In these materials the lack of compromise of the interacting magnetic degrees of freedom with a long range order may be due to a plethora of collective low energy configurations offered by lattices that often have triangular motifs and/or low connectivity \cite{lee2002emergent,moessner2006geometrical}. Prototypical examples are spin glasses and spin ice materials \cite{PhysRevLett.64.2070, castelnovo2012spin, ramirez1999zero}.  In the case of spin glasses,  frustration is derived from bond disorder \cite{PhysRevB.16.4630,lacroix2011introduction}.
In this case no  long-range order of ferromagnetic
or antiferromagnetic type can be established. Instead, the materials freeze into a state where the spins are aligned in random directions \cite{binder1986spin} and magnetic correlations cancel out. Therefore, the understanding of the glass transition into the freezing state in these materials relies in the examination of their dynamics \cite{keren1996probing}. In this respect a key aspect of spin glass research lies in the study of the time autocorrelation function.

In spin ice materials the dipolar interactions and the  weak antiferromagnetic superexchange realized in rare-earth pyrochlores result in an effective ferromagnetic coupling that in combination with single site anisotropy, yield a frustrated spin arrangement that mimics the geometric frustration in water ice \cite{snyder2001spin}.  Here,  the study of the thermal relaxation processes by means of a.c. magnetic susceptibility measurements \cite{castelnovo2012spin,ehlers2004evidence} has revealed a magnetic monopole like dynamics mediated by the Coulomb interaction between charges \cite{ryzhkin2013dynamic}. 

Modern lithographic techniques have allowed the fabrication and study of the artificial counterpart of spin ice in two dimensions \cite{wang2006artificial,drisko2015fepd} the so called  artificial spin ices, ASI \cite{nisoli2013colloquium}. 
 \begin{figure}
\includegraphics[width=\columnwidth]{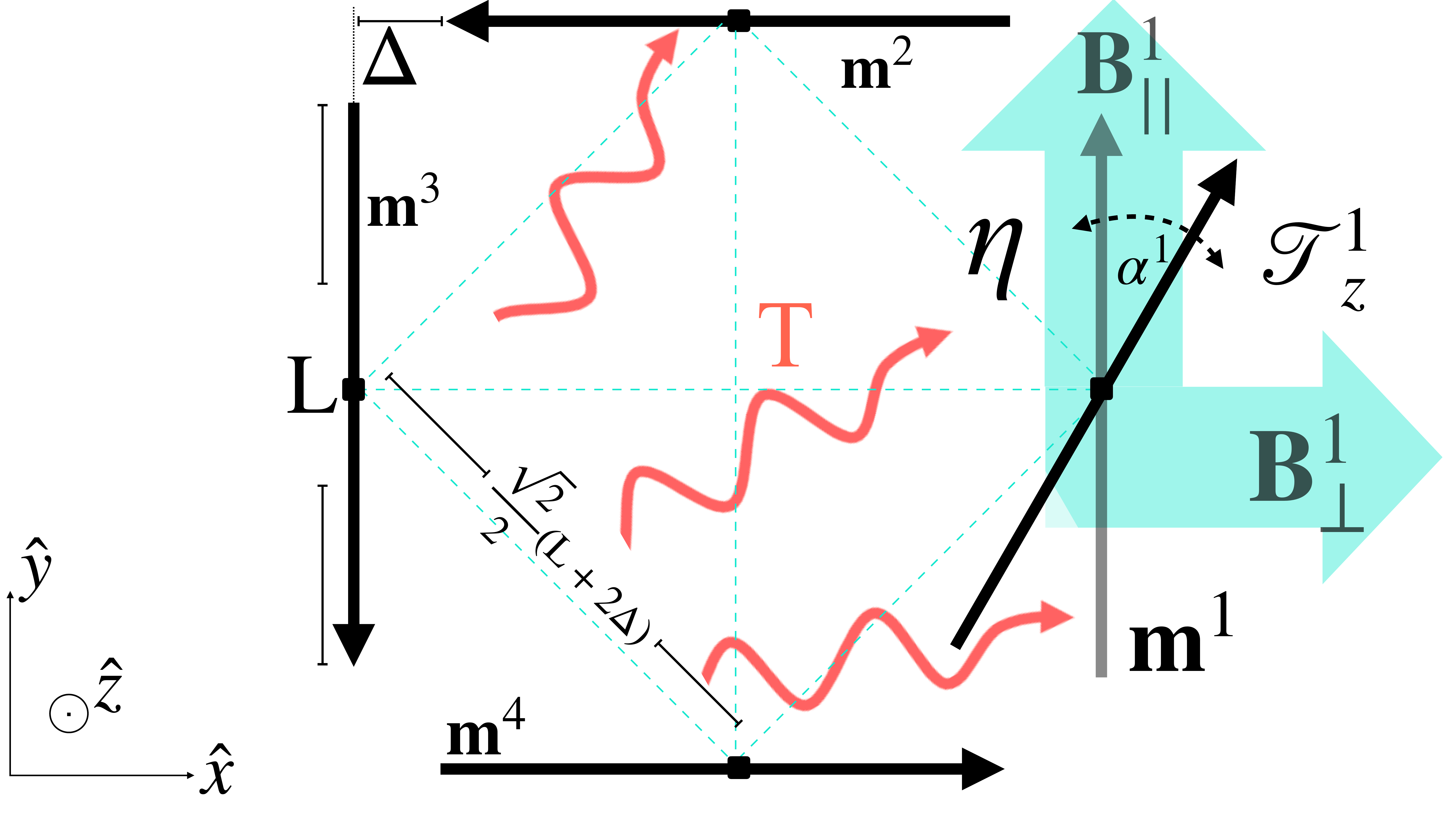}
\caption{(color online) Dipolar square plaquette of lattice constant $\frac{\sqrt{2}}{2}(L+2\Delta)$ in the vortex configuration. The dipoles with magnetic moment $\bm{m}^i$,  length L and moment of inertia I, are represented by black arrows. They rotate with angle $\alpha^i$ in the $x-y$ plane. The system is under finite temperature T and the viscous rotation of the magnets is illustrated by the parameter $\eta$. Dotted (cyan) lines joining dipoles illustrate the magnetic dipolar interaction between them. This interaction gives rise to the magnetic field $\bm{B}^1=(B^1_{||},B_{\perp}^1)$ at the position of the dipole $\bm{m}^1$ yielding the magnetic torque $\mathcal{T}^1_z$ responsible of the rotation of $\bm{m}^1$ respect to the $\hat{z}$ axis.}
\label{fig:f1}
\end{figure}
They are a subset of artificial dipolar systems \cite{leo2018collective} which have become ideal settings for observing dynamical effects in magnetic systems. In artificial spin ice structures \cite{kapaklis2012melting}, the arrangement of moments product of elongated single-domain nanopatterned magnetic islands can lead to excited states with magnetic charges \cite{mellado2010dynamics}, analogous to the monopole excitations reported in rare-earth pyrochlores  \cite{mengotti2011real}. Recently, susceptibility measurements \cite{pohlit2020collective} of thermally active extended square ASI  \cite{farhan2013direct} revealed that magnetic fluctuations and excitation population depend on lattice spacing and interaction strength between islands \cite{kapaklis2014thermal}. 
With the purpose of extracting parameters related to the magnetostatic energies of ASI arrays directly from the susceptibility measurements \cite{pohlit2020collective} a Vogel-Fulcher-Tammann  law \cite{garca1989theoretical}, has recently been employed. Nevertheless the results showed that this approach fails to address the dynamics of thermal ASI arrays. The failure of this and other phenomenological models for describing the dynamic response from frequency measurements in systems as diverse as spin ices, spins glasses and superconductors \cite{rault2000origin, sankar2018vortex} is rooted in the ad-hoc time scale distributions used to complement models originated from Debye processes. 

In this paper we present a prototype model that illustrates a different approach aimed to unveil the specific role played by each of the constituents that characterize a dipolar array in the stages of dynamical evolution. The model consists of square plaquettes made out of interacting inertial dipoles (or dipolar needles) which rotate in a viscous media in the $x-y$ plane, see Fig.~\ref{fig:f1} and Figs.~\ref{fig:f3}(a,b).
The systems dynamics is modeled by a Langevin equation with gaussian thermal noise \cite{ullersma1966exactly} and dipolar interactions. The analytical solution of the Langevin equation for small angular oscillations allows to identify the relevant time scales for the thermal relaxation dynamics and detect their manifestation in the time autocorrelation function $\mathcal{C}(s)$. We found that the systems proper frequencies originate from the interplay between the internal magnetic field due to dipolar interactions, temperature and intrinsic features such as inertia and damping. Further analysis of $\mathcal{C}(s)$  allows to exhibit the qualitative differences in the dynamical response of edge and bulk states in magnetic arrays. The approximated solution of $\mathcal{C}(s)$, valid for short times, is corroborated and complemented by molecular dynamics simulations. The numerical approach allows the study of the magnetization loops of edge and bulk states when external magnetic fields are applied. Here the anisotropy of the dipolar interactions sustained by dipoles located at the edge and the bulk of the lattices manifests as a magnetization plateaux at $m=1/3$.

The paper is organized as follows: in Section \ref{sec:model} we give an overview of the magnetic dipolar energy to account for the interactions between magnets in our system. Then the Langevin equation is introduced to account for the thermal dynamics of the dipoles and at the end, we give a brief summary of the molecular dynamics simulations employed in the paper.  Section \ref{sec:Results} is devoted to the results. In the first part we study the system proper time scales obtained from the equations of motion. Then we derive and examine the time autocorrelation function in the non interacting limit by considering the case of an isolated dipole. Next we address the dynamics of a set of interacting dipoles by deriving the time autocorrelation function of a square plaquette and identifying the stages of relaxation in terms of the system proper time scales. The following subsection generalizes the previous case to a cluster made out of four plaquettes with the purpose of comparing the  thermal relaxation of the edges and the bulk of dipolar arrays. We end Section \ref{sec:Results} by addressing the magnetization dynamics of square clusters.  In
the Conclusion \ref{sec:Conclusions} we summarize our findings. Technical details are given in the Appendix at the end of the paper. 
\section{Model}
\label{sec:model}
\subsection{Interaction between magnets}
\label{sec:Interaction between magnetic needles}
The system consists of a set of $x-y$ interacting dipoles of length $L$, mass $\it{m}$ and moment of inertia  $I=\frac{\it{m}L^2}{12}$. The magnets are located at the vertices of square plaquettes where the distance between the centers of two nearest neighbor dipoles is $\frac{\sqrt{2}}{2}(L+2\Delta)$ as shown in Fig.~\ref{fig:f1}. The position of the center of dipole $i$ is denoted ${\bm r}^i$ and the director vector joining two dipoles is given by $\hat {\bm{e}}^{ik}= \frac{({\bm r}^i -{\bm r}^k )}{|{\bm r}^i -{\bm r}^k |}$. The rotation of a dipole occurs in the $x-y$ plane and is described in terms of the angle $\alpha^i$ chosen with respect to its equilibrium position.  This rotation is viscous and the damping parameter is denoted by $\eta$. Here the magnetic moment of a magnet of radius $r$ and saturation magnetization $M_s$ is ${\bm m}^i = m_0   \hat{{\bm m}}^i $. The unit vector $\hat{{\bm m}}^i  = (\cos\alpha^i,\sin\alpha^i)$ and the magnetic moment intensity $m_0=qL$ $(\rm[m^2\,A])$, where $q$ represents a magnetic charge defined as $q=\pi r^2 M_s $  \cite{mellado2012macroscopic}.  
\subsubsection{Dipolar coupling} 
The magnetic dipoles interact by means of the magnetic dipolar energy as follows:
\begin{eqnarray}
\mathcal{U}_{dip}=\frac{\gamma}{2} \sum_{i\neq k=1}^n \frac{\hat {\bm  m}^i \cdot\hat {\bm m}^k - 
3 (\hat {\bm m}^i \cdot \hat {\bm{e}}^{ik} )(\hat {\bm m}^k\cdot \hat {\bm{e}}^{ik} )}{|{\bm r}^i -{\bm r}^k |^3},
\label{eq:Energy}
\end{eqnarray}
 where $\gamma =\frac{\mu_0\,m_0^2}{4\pi}$ ($[\rm N\,m^4]$) and $\mu_0$ is the magnetic permeability in vacuum.  In Eq.~(\ref{eq:Energy})  the geometrical parameter $\Delta$ hidden in ${\bm r}$ changes the distance between magnets and therefore it tunes the strength of the dipolar coupling (Fig.~\ref{fig:f1}). A set of n dipoles gives rise to a magnetic field at the position of dipole $\bm{m}^i$, which has the form  
\begin{eqnarray}
\bm{B}^i=-\frac{\mu_0m_0}{8\pi} \sum_{k\neq i=1}^n \frac{\hat {\bm m}^k - 
3 \hat {\bm{e}}^{ik} (\hat {\bm m}^k\cdot \hat {\bm{e}}^{ik} )}{|{\bm r}^i -{\bm r}^k |^3}
 \label{eq:field}
\end{eqnarray}
this magnetic field yields a torque on $\bm{m}^i$ given by $\mathcal{T}^i_z=(\bm{m}^i\times \bm{B}^i)_z=m_x^i B^i_y-m_y^i B^i_x$ which rotates $\bm{m}^i$ around the $\hat{z}$ axis as illustrated in Fig.~\ref{fig:f1}.
\subsection{Thermal dynamics}
\label{sec:Thermal dynamics of the magnetic needles}
Here we address the viscous dynamics of interacting magnetic dipoles at finite temperature.  For that effect we study the equation of motion for the angular rotation of the dipoles in contact with a heat bath. The forces that determine the torques and thus the rotation of each dipole are 1) the dipolar forces product of the dipolar interaction between them, 2) a frictional force due to the viscous rotation of the inertial magnets and 3) a random force $\xi(t)$ which accounts for thermal fluctuations. Therefore a dipole $k$ will be modeled as a Brownian particle in the potential $\rm V_{dip}^k=\mathcal{U}_{dip}^k/m^k$ and in contact with a thermal bath of temperature T. Consequently, its dynamics will be governed by the Langevin equation as shown next \cite{sekimoto1998langevin}. 
\subsubsection{Langevin Equation}
The dynamics of the angular variable $\alpha^i$ is described by the Langevin equation \cite{sekimoto1998langevin}: 
\begin{eqnarray}
I\frac{d^2\alpha^i}{dt^2}=\sqrt{2\eta k_B T}\xi(t)-\mathcal{\eta}\frac{d\alpha^i}{dt}-\mathcal{T}^i_z,
\label{eq:langevin1}
\end{eqnarray}
where $I$ ([Kg $\rm m^2$]) is the inertia moment of each dipole, and $\eta$ ([$\frac{\rm Kg\,m^2}{s}$]) is the damping coefficient that accounts for its viscous rotation. Thermal fluctuations due to the coupling of the magnet with the thermal bath are modeled by a $\delta$-correlated Gaussian noise $\xi(t)$ of zero mean and unit intensity: $\rm\langle \xi(t)\rangle=0,$ $\rm\langle \xi(t) \xi(t')\rangle=\delta(t-t')$. The units of $\xi(t)$ are $[1/\sqrt{s}]$ and $k_B$ is the Boltzmann constant. The term $\mathcal{T}^i_z=(\bm{m}^i\times \bm{B}^i)_z$ accounts for the magnetic torque along the $\bm z$ direction on dipole $\bm{m}^i$ due to the net internal magnetic field originated by all other dipoles in the system.  Such a torque is $\mathcal{T}^i_z=(\bm{m}^{i}\times \bm{B}^i)_z=m_0(B_{\bot}^{i}\cos\alpha^i -B_{||}^{i}\sin\alpha^i)$, where $B_{\bot}^{i}$ and $\rm B_{||}$ are respectively the fields perpendicular and parallel to the direction of $\bm{m}^i$ at equilibrium. Considering the simplified case where 1) dipoles deviate slightly from their equilibrium positions and 2) at a given position the total internal fields are such that $\rm |B_{\bot}^i|\ll |B_{||}^i|$ (this assumption will be justified in Section \ref{sec:Results}), yields $\rm \mathcal{T}^i_z\sim m_0 B_{||}^i\alpha^i\equiv\mathcal{K}^i\alpha^i$. Under these circumstances, Eq.~(\ref{eq:langevin1}) becomes, 
\begin{eqnarray}
I\frac{d^2\alpha_i}{dt^2}=\sqrt{2\eta k_B T}\xi(t)-\mathcal{\eta}\frac{d\alpha^i}{dt}-\mathcal{K}^i\alpha^i,
\label{eq:langevin2}
\end{eqnarray}
  The simplified version of the Langevin equation,  Eq.~(\ref{eq:langevin2}) is used to compute the analytical results presented along the paper, while the full version of the Langevin equation, Eq.~(\ref{eq:langevin1}) is used to address the problem using numerical simulations.
 \subsection{Molecular Dynamics Simulations}
\label{sec:Molecular Dynamics simulations}
Numerical results were obtained by direct numerical integration of the equations of motion, Eq.~(\ref{eq:langevin1}), for each dipole interacting with all the others via dipolar interactions. We used a Verlet method with an integration time step $\Delta t=2\times10^{-6}$.  To produce $\xi(t)$ a given temperature T was multiplied by a random number with a gaussian distribution. In all simulations the same parameters for lattice constant, damping, inertia and magnitude of the magnetic moments of the dipoles were used (see Appendix \ref{app:Molecular dynamics simulations of the Langevin dynamics} for details), otherwise stated. 
\section{Results}
\label{sec:Results}
\subsection{Time scales}
\label{sec:Time scales}
Eq.~(\ref{eq:langevin2}) allows to identify four meaningful times scales that determine the thermal dynamics of the dipoles in the system. 
The relaxation time of the angular velocity from the inertial and damping contributions sets the proper time scale $\tau_1\equiv\frac{I}{\eta}$. The angular relaxation time from the damping and the internal magnetic field set $\tau_2\equiv\frac{\eta}{\mathcal{K}}$. 
The time scale given by the rate between inertia (which depends on the length and mass of the magnetic degrees of freedom) and the internal dipolar fields set $\rm \tau_3\equiv\sqrt{\frac{I}{\mathcal{K}}}$. Finally the proper time $\tau_{\rm th}\equiv\rm\frac{\eta}{k_B T}$ weighs thermal up to damping energies. Here, we found that the minimum time scale is set by $\tau_3$. Notice that while $\rm\tau_1$ and $\rm\tau_{th}$ are related to the system single particle aspects, $\tau_2$ and $\tau_3$ arise due to the interaction between dipoles.   Since $\mathcal{K}\sim\rm\frac{\mu_0m_0^2}{\Delta^3}$ ($\Delta$ tune the distance between dipoles in the array) then $\tau_2\rm\sim \frac{\eta \Delta^3}{\mu_0 m_0^2}$ and $\tau_3\rm\sim(\frac{I}{\mu_0 m_0^2})^{1/2}\Delta^{3/2}$ which demonstrates that the proper time scales $\tau_2$ and $\tau_3$ decrease for larger lattice constants. 

The above identified proper frequencies are useful to express Eq.~(\ref{eq:langevin2})  in its dimensionless form:
\begin{eqnarray}
\frac{d^2\alpha^i}{ds^2}=\sqrt{2 \nu}\tilde{\xi}(s)-\frac{d\alpha^i}{ds}-\frac{\tau_1}{\tau_2}\alpha^i,
\label{eq:langevin3}
\end{eqnarray}
where $\rm s\equiv\frac{t}{\tau_1}$ has become now the dimensionless time and $\rm \frac{\tau_1}{\tau_2}=\frac{I \mathcal{K}}{\eta^2}$. The rescaled gaussian noise has the same statistics as $\xi(t)$ but now $\tilde{\xi}(s)$ has no units.  $\nu\equiv\frac{\rm\tau_1}{\rm\tau_{th}}=\frac{I k_B T}{\eta^2}$ is the rescaled thermal noise.  

If the system time scales are such that $\tau_1\ll \tau_2$, the last term in the right hand of Eq.~(\ref{eq:langevin3}) can be neglected. Furthermore, if  $\rm\tau_1\ll \tau_2$ then $\rm\frac{I\mu_0m_0^2}{\eta^2}\ll \Delta^3$. Therefore weighting the lattice constant with respect to the dipoles intrinsic properties becomes a suitable criterion to estimate whether a dipolar array  behaves as a weakly interacting system (a strongly damped or diluted) or a strongly interacting one. 
\subsection{Non interacting limit} 
\label{sec:Non interacting limit.}
We begin by examining the case of a dilute array of magnets. As a non interacting limit consider the thermal relaxation of an isolated dipole right after an initial weak perturbation has taken it away from equilibrium. Its dimensionless Langevin equation reads: $\frac{d^2\alpha}{ds^2}=\sqrt{2\nu}\tilde{\xi}(s)-\frac{d\alpha}{ds}
$ that corresponds to an Ornstein-Uhlenbeck process \cite{bojdecki1991gaussian} with mean square angular rotation $
\langle\delta\alpha(s)^2\rangle=\nu\big(s-1+e^{-s}\big)
$, where $\delta\alpha(s)=(\alpha(s)-\alpha(0))$. 
The thermal relaxation of the dipole can be captured through its time autocorrelation function, 
 \begin{eqnarray}
\mathcal{C}(s)=\langle\bm{\hat{m}}(s)\cdot\bm{\hat{m}}(0)\rangle=\operatorname{Re}\langle e^{i\delta\alpha(s)}\rangle,
\label{eq:}
\end{eqnarray}
\begin{figure*}
\includegraphics[width=\textwidth]{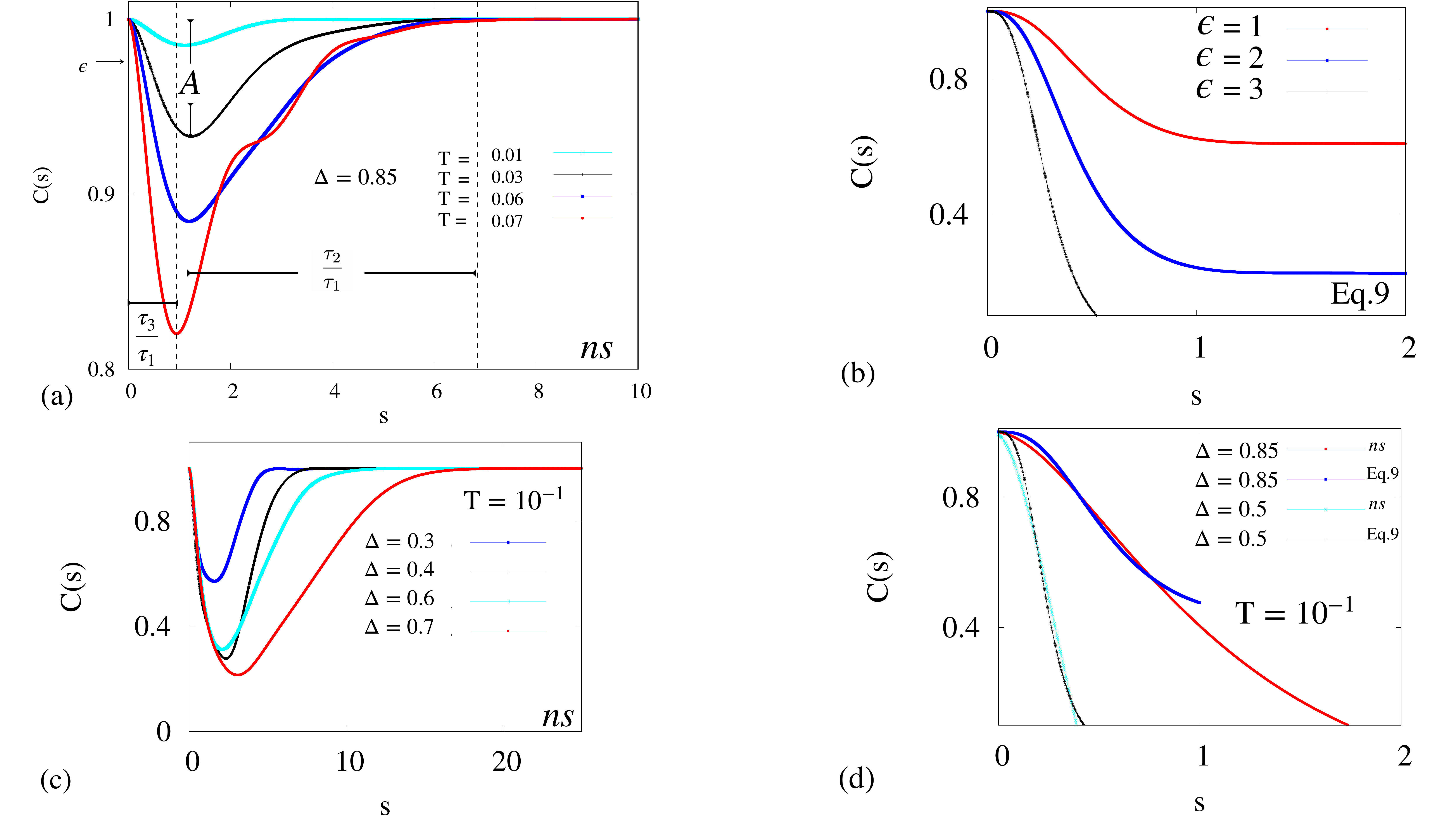}
\caption{(color online) $\mathcal{C}(s)$ of the square plaquette  relaxing into the vortex configuration shown in Fig.~\ref{fig:f1}. $s=\frac{t}{\tau_1}$ is the dimensionless time. (a) Numerical evaluation (ns) of $\mathcal{C}(s)$ with $\Delta=0.85$ (in units of $\ell$). Different curves correspond to different values of $T$ (in units of $\mathcal{K}(L)$). The different stages of the evolution of $\mathcal{C}(s)$ are associated to the system time and energy scales as explained in the text. $A$ denotes the amplitude of the early time oscillations. (b) evaluation of  Eq.~(\ref{eq:auto}) comparing the early evolution of $\mathcal{C}(s)$ for three values of $\epsilon$. (c) Likewise (a), but now $T=10^{-1}$ and different curves correspond to $\mathcal{C}(s)$ computed at different $\Delta$. 
 (d) comparison of the numerical solution of $\mathcal{C}(s)$ (in red and cyan) with Eq.~(\ref{eq:auto}) (in blue and black) during the early stage of thermal relaxation for two values of $\Delta$ and $T=10^{-1}$.}
\label{fig:f2}
\end{figure*}
Because  $\delta\alpha(s)$ is linear in the noise and $\tilde{\xi}(s)$ has a Gaussian distribution,  $\delta\alpha(s)$ is also Gaussian with a zero mean and a second moment $\langle(\delta\alpha(s))^2\rangle$ \cite{pathria2011statistical}. For a gaussian variable $x$, with a mean $\mu_x$,  and a variance $\sigma_x$,
$\langle e^{iA}\rangle=e^{iA\mu_x-\frac{A^2}{2}\sigma_x^2}$ and therefore
 \begin{eqnarray}
 \mathcal{C}(s)=e^{-\frac{\langle{(\delta\alpha(s)})^2\rangle}{2}}
  \end{eqnarray}
for a single dipole it yields:
 \begin{eqnarray}
 \mathcal{C}^{1}(s)=e^{-\nu\big(s-1+e^{-s}\big)}
  \end{eqnarray}
which depends on the rescaled thermal noise only (see Appendix \ref{app:Solution of Langevin equation} for details). For short times $t \ll \tau_1$, $\mathcal{C}^{(1)}(s)\sim e^{-\nu \frac{s^2}{2}}$,  while for long times $t \gg \tau_1$, $\mathcal{C}^{(1)}(s)\sim e^{-\nu s}$. This shows that in a non interacting array, the thermal relaxation may be slowed down by decreasing the temperature or by increasing the damping to inertia quotient of the magnetic degrees of freedom in the system.
 \begin{figure*}
\includegraphics[width=\textwidth]{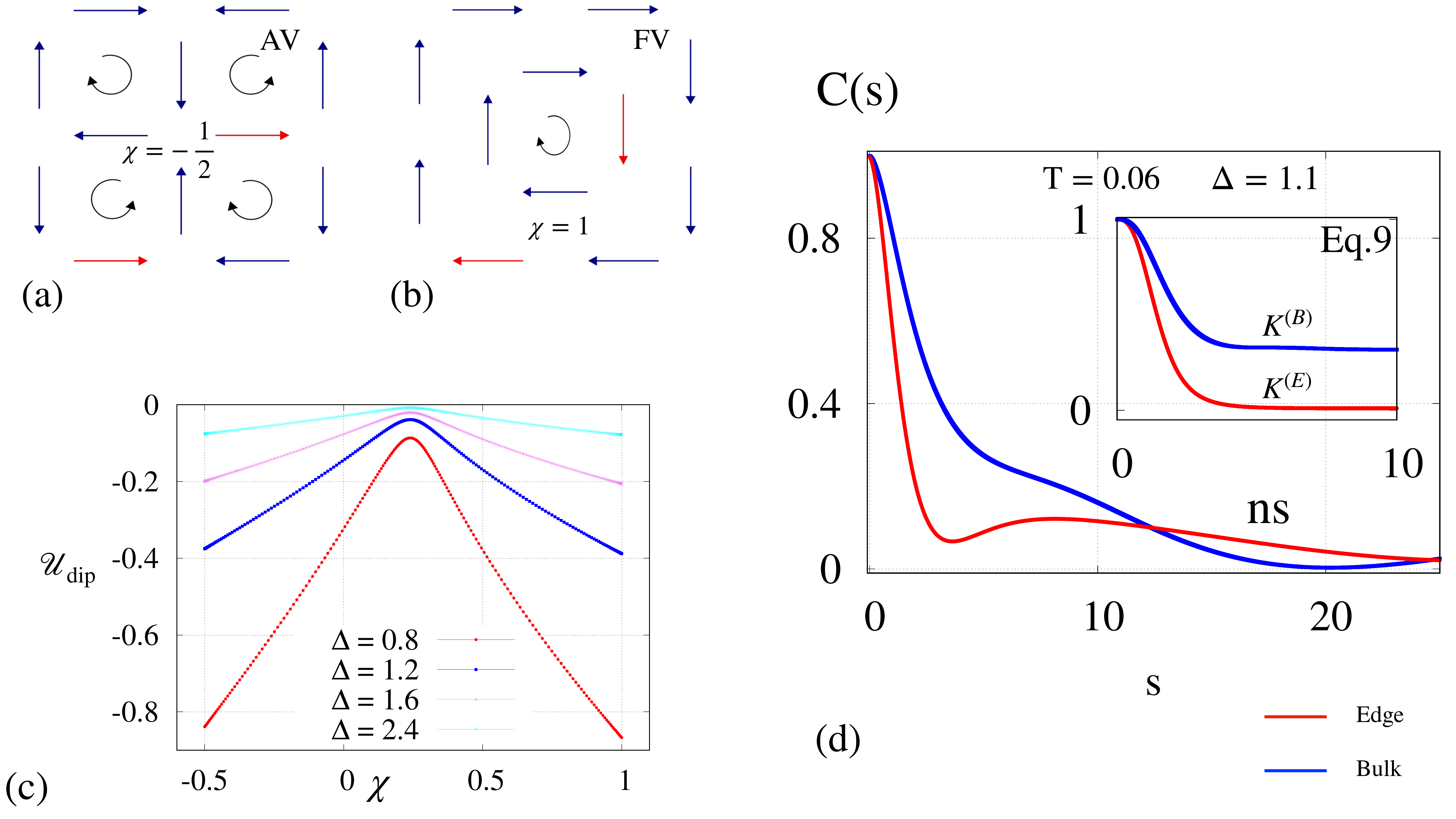}
\caption{(color online)  (a) Antiferromagnetic vortex state ($\rm AV$) and (b) ferromagnetic vortex state ($\rm FV$) for a small cluster made out of four square plaquettes after relaxation. In (a) and (b) the red dipoles highlight  edge and bulk magnets examined in the text.  (c) dipolar energy density (in units of $\mathcal{K}(L)$) of the square cluster versus its chirality for several values of $\Delta$ (d) numerical results comparing $\mathcal{C}(s)$ of edge (in red) and bulk (in blue) dipoles at the sites of the cluster in the $\rm FV$ state with $\Delta=1.1$ and at $\rm T=0.06$. The inset shows evaluation of $\mathcal{C}(s)$ from Eq.~(\ref{eq:auto}) of edge (using $\mathcal{K}^{(E)}$) and bulk  dipoles (using $\mathcal{K}^{(B)}$) with $\Delta=1.1$ and $\rm T=0.06$.}
\label{fig:f3}
\end{figure*}
\subsection{Interacting case} 
\label{sec:Interacting case}
Next we proceed to study the thermal relaxation in the case of a set of interacting dipoles.  Numerical solution of Eq.~(\ref{eq:langevin1}) for temperatures $T\ll \mathcal{K}$ (and energy minimization at T=0) returned the square plaquette settled into the magnetic vortex configuration  shown in Fig.~\ref{fig:f1} (or its time reversal). 
\subsubsection{Time autocorrelation function $\mathcal{C}(s)$} 
Consider, the dipole with magnetic moment $\bm{m}^1$ in Fig.~\ref{fig:f1}.  The torque sustained by $\bm{m}^1$ along the $\bm{z}$ direction  due to the three other magnets is $(\bm{m}^1\times\bm{B}^1)_z=m_0(B_{\bot}^1\cos(\beta^1+\alpha^1) -B_{||}^1\sin(\beta^1+\alpha^1))$, where $\beta^1$ is its equilibrium angle and $\alpha^1$ is a small angular deviation. In the vortex magnetic configuration of Fig.~\ref{fig:f1},  $\beta^1=n\pi$ (n integer), $B_{\bot}^1$ cancels out and $|B_{||}^1|=\frac{\text{m}_0 \mu _0\Lambda}{(L+2\Delta )^3}$ with $\Lambda\equiv\left(1+6 \sqrt{2}\right)$ a geometrical factor due to the point symmetry of the sites forming the lattice. The square plaquette has four oscillation modes. In the lowest energy mode, parallel dipoles oscillate in phase and small deviations out of the equilibrium barely change $\Lambda$. Thus $\sin{(\beta^1+\alpha^1)}\sim \alpha^1$ along with $\frac{B_{\bot}^1}{B_{||}^1}(\alpha^1)\rightarrow 0$ produce that at the mean field level $|(\bm{m}^1\times\bm{B}^1)_z|\sim \frac{\mu _0\text{m}_0^2 \Lambda}{(L+2\Delta )^3}\alpha^1=\mathcal{K}\alpha^1$. Symmetry ensures that $\mathcal{K}=\frac{\mu _0\text{m}_0^2 \Lambda}{(L+2\Delta )^3}$ is equivalent for all dipoles at the square plaquette.

With this approximation the Langevin Eq.~(\ref{eq:langevin2}) can be solved by constructing the green function $\mathcal{G}$ that verifies $I\ddot{\mathcal{G}} +\eta\dot{\mathcal{G}}+\mathcal{K}\mathcal{G}=\delta(t-t^{'})$ as shown in Appendix \ref{app:Solution of Langevin equation}. Indeed, we can use $\mathcal{G}$ to find the mean square angular oscillations of the interacting dipoles in the case of small angular deviations:
 \begin{eqnarray}
 \langle \delta\alpha(s)^2\rangle&=&\epsilon-\frac{\epsilon}{\zeta^2}\left[1-\cos{(s\zeta)}+\zeta\sin{(s\zeta)}+\zeta^2\right]e^{-s}
\label{eq:msrsquare}\nonumber
\end{eqnarray}
where $\epsilon\equiv\rm\frac{\tau_2}{\tau_{th}}=\frac{\Delta^3 k_B T}{\mu_0m_0^2}$ weights the thermal to the dipolar energy and $\zeta\equiv\sqrt{4\frac{\tau_1}{\tau_2}-1}=\sqrt{\frac{\mu_0m_0^2I}{\eta^2\Delta^3}}$ rates the geometrical and magnetic aspects of the dipoles to the damping and interactions in the lattice.  For  $\tau_1\geq\frac{\tau_2}{4}$ (or $\rm I\geq\frac{\eta^2}{4\mathcal{K}}$),  and for short times $t \leq \tau_3$,  the time autocorrelation function becomes,
 \begin{eqnarray}
 \mathcal{C}(s)=e^{\epsilon-\frac{\epsilon}{\zeta^2}\left[1-\cos{(s\zeta)}+\zeta\sin{(s\zeta)}+\zeta^2\right]e^{-s}}
\label{eq:auto}
\end{eqnarray}
In the limit of weak interactions Eq.~(\ref{eq:auto}) yields $\mathcal{C}(s)=e^{-\frac{\epsilon}{2}s^2}=\mathcal{C}^{(1)}(s)$,  the autocorrelation of an isolated dipole.  \\
\subsubsection{Geometrical factor $\Lambda$ and correlation length $\ell$} 
As mentioned above, when the condition $\tau_1=\tau_2$ is met, the dynamics of the plaquette changes from a weakly to a strongly interacting regime. At zero temperature, this transition occurs when the lattice constant is such that $L+2\Delta=\Lambda^{1/3}\ell$. $\Lambda$ contains information about the symmetry of the lattice (for instance for a triangular plaquette it changes to $\Lambda=(3-\sqrt{3})\sqrt{2}$).  $\ell=(\frac{\rm \mu_0 m_0^2  I}{\eta^2})^{1/3}$ on the other side, sets a new length scale that depends on the dipoles intrinsic properties only.  Furthermore, $\ell$  determines a magnetic correlation length on account of the intrinsic properties of the magnetic degrees of freedom, such as inertia, damping and the intensity of their magnetic moments. Therefore, an array can be categorized in the strongly correlated regime when $\Delta\ll\ell$.
While damping contributes to reduce $\ell$,  inertial effects increase the correlation length between magnets, which is also enhanced by increasing the intensity of their magnetic moments. For the square plaquette $\tau_1=\tau_2$ for $\ell^*=\frac{L+2\Delta}{\left(1+6 \sqrt{2}\right)^{\frac{1}{3}}}$. 
Henceforth, we normalize all length scales by $\ell$, otherwise stated. 
\subsubsection{$\mathcal{C}(s)$ versus T, $\Delta$ and $\epsilon$.} 
In what follows we study the evolution of the time autocorrelation function of the dipoles in the square plaquette in terms of the temperature of the system, the strength of the dipolar interactions and the rate between thermal to dipolar couplings. To complement the results obtained from Eq.~(\ref{eq:auto}) which are valid for short times ($t \leq\tau_3$) and small angular oscillations, we have run molecular dynamics simulations where  T,  the strength of the dipolar interactions set by $\Delta$ and $\epsilon$ have been varied.  $I$, $\eta$ and $m_0$ on the other hand stayed fixed. Details can be found in Appendix \ref{app:Molecular dynamics simulations of the Langevin dynamics}.

Hereinafter temperature T is measured in units of the magnetic energy between two nearest dipoles located in the square plaquette with $\Delta=L$, $\mathcal{K}(L)$, and $\Delta$ is measured in units of the correlation length $\ell$. 

Next, we discuss Fig.~\ref{fig:f2} respect to aspects such as the onset of the relaxation, the amplitude $A$ of the oscillations of $\mathcal{C}(s)$ at short times, and the qualitative different stages of its dynamical evolution in terms of the intrinsic features of the magnetic degrees of freedom and the geometrical aspects of the lattice. Fig.~\ref{fig:f2}(a), shows the numerical solution of $\mathcal{C}(s)$ belonging to a square plaquette that relaxes from a slightly perturbed state (from its equilibrium configuration Fig.~\ref{fig:f1}) at several values of T for the case of fixed interactions ($\Delta=0.85$).  At the onset of the thermal relaxation ($t<\tau_3$)  we observe that the autocorrelations computed at larger values of T (red and blue curves) decay earlier from $1$ than the others. In addition, when examining Eq.~(\ref{eq:auto}) and the formula for the angular deviations,  it is apparent that the amplitude $A$ of the oscillations of $\mathcal{C}(s)$ is controlled by $\epsilon$, the ratio between thermal and dipolar interactions. Indeed $\epsilon$ scales like $\sim T\Delta^3$, therefore when $\Delta$ is fixed as in Fig.~\ref{fig:f2}(a), larger temperatures increase $\epsilon$ and decrease the time scale $\tau_{th}$ triggering an earlier departure of the system from its initial magnetic configuration. Fig.~\ref{fig:f2}(c), shows the effect of $\Delta$ for fixed values of temperature ($\rm T=10^{-1}$).  Here with smaller interactions (larger $\Delta$) $\epsilon$ grows and therefore $A$ becomes larger. The growing of $\Delta$ also has the effect of increasing $\tau_3$ and delaying the turning point of $\mathcal{C}(s)$. Further, a reduction of the dipolar coupling amplifies the relative effect of the inertia of the magnets, which explains the increment in the size of the oscillations in $\mathcal{C}(s)$. It has also the effect of reducing the effective stiffness of the system due to dipolar interactions \cite{mellado2012macroscopic} which means that it takes longer for the magnets to return to its equilibrium state as illustrated by the increment of $\tau_2$ with $\Delta$ as shown in Fig.~\ref{fig:f2}(c). This behavior is also captured by the analytical counterpart Eq.~(\ref{eq:auto}), as is manifested in Fig.~\ref{fig:f2}(b)) that shows a consistent change of $\mathcal{C}(s)$ with  $\epsilon$. Figs.\ref{fig:f4A}(a) and \ref{fig:f4A}(b) in Appendix \ref{app:amplitude} show $A$ with respect to  T and $\Delta^3$ respectively consistent with this analysis.  

As mentioned above, the evolution before $\mathcal{C}(s)$ has reached its minimum value, is controlled by inertia and interactions and lasts $t\sim\tau_3$. Fig.~\ref{fig:f2}(a))  shows that the minimum of $\mathcal{C}(s)$ is reached at roughly the same s for all curves since $\Delta$  and therefore $\tau_3$ remain constant.  Lastly, the compromise between damping and interactions carries the system back to the equilibrium vortex configuration after a time $t\sim\tau_2$ has elapsed in all cases. 

In Fig.~\ref{fig:f2}(d) we compare the numerical solution of $\mathcal{C}(s)$ and Eq.~(\ref{eq:auto}) for short times, ($t\leq\tau_3$), for two values of $\Delta$ and T=$10^{-1}$, confirming the agreement between Eq.~(\ref{eq:auto})  and the numerical solution at the early stage of thermal relaxation.
 \begin{figure}
\includegraphics[width=\columnwidth]{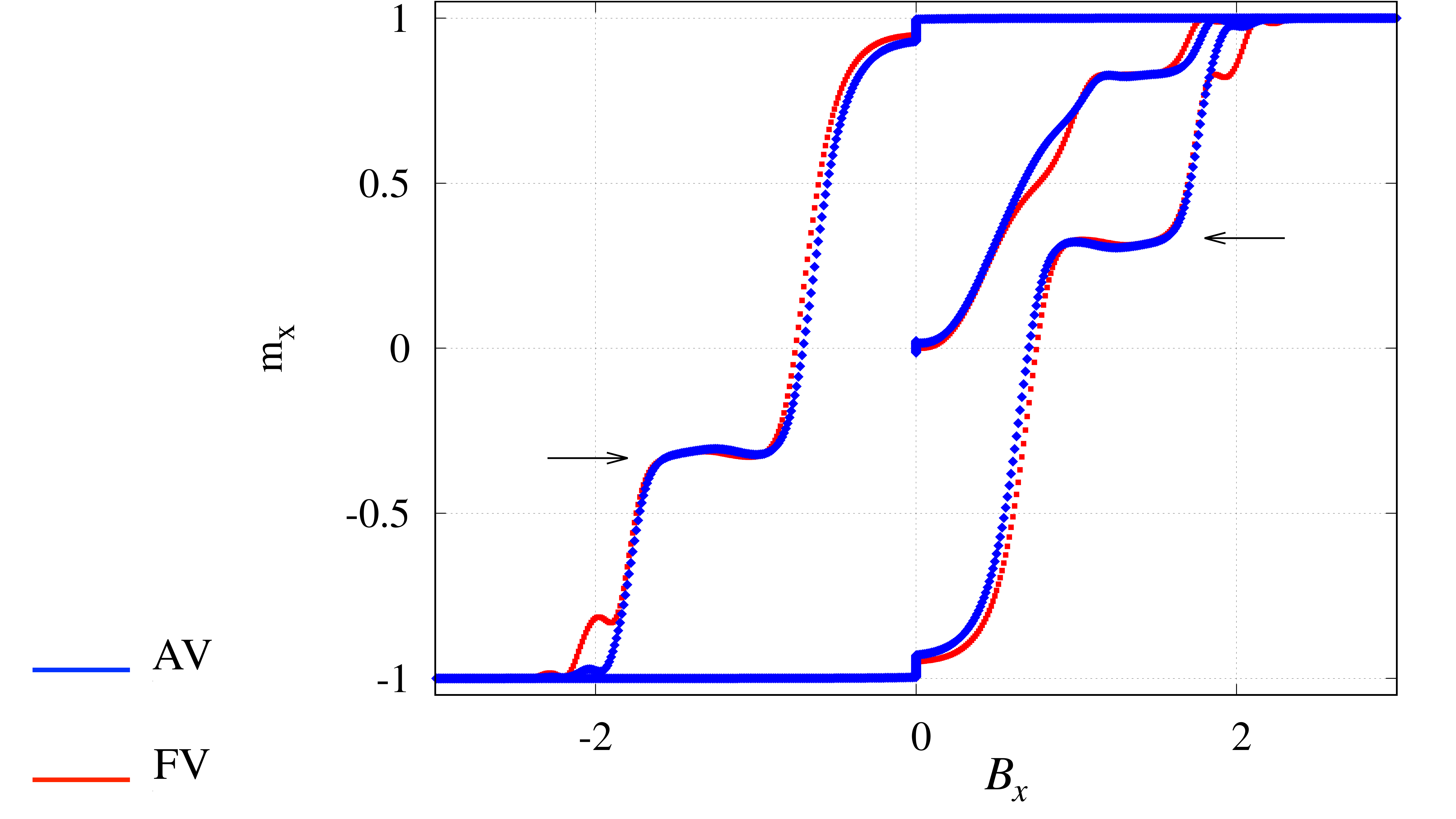}
\caption{(color online) Magnetization parallel to the applied field direction. Blue curve shows the result for the cluster in the $\rm AV$ (Fig.~\ref{fig:f3}(a)) with $\Delta=1.1$ and $\rm T=6\times 10^{-2}$ while the red curve shows the results of the lattice in the $\rm FV$ (Fig.~\ref{fig:f3}(b)) with $\Delta=1$ and $\rm T=2.2\times 10^{-2}$. Magnetization $\rm m_x$ is in $\rm m_0$ units and the magnetic field $B_x$ is in units of $\frac{\mathcal{K}(\Delta)}{m_0}$.}
\label{fig:f4}
\end{figure}
 \begin{figure}
\includegraphics[width=\columnwidth]{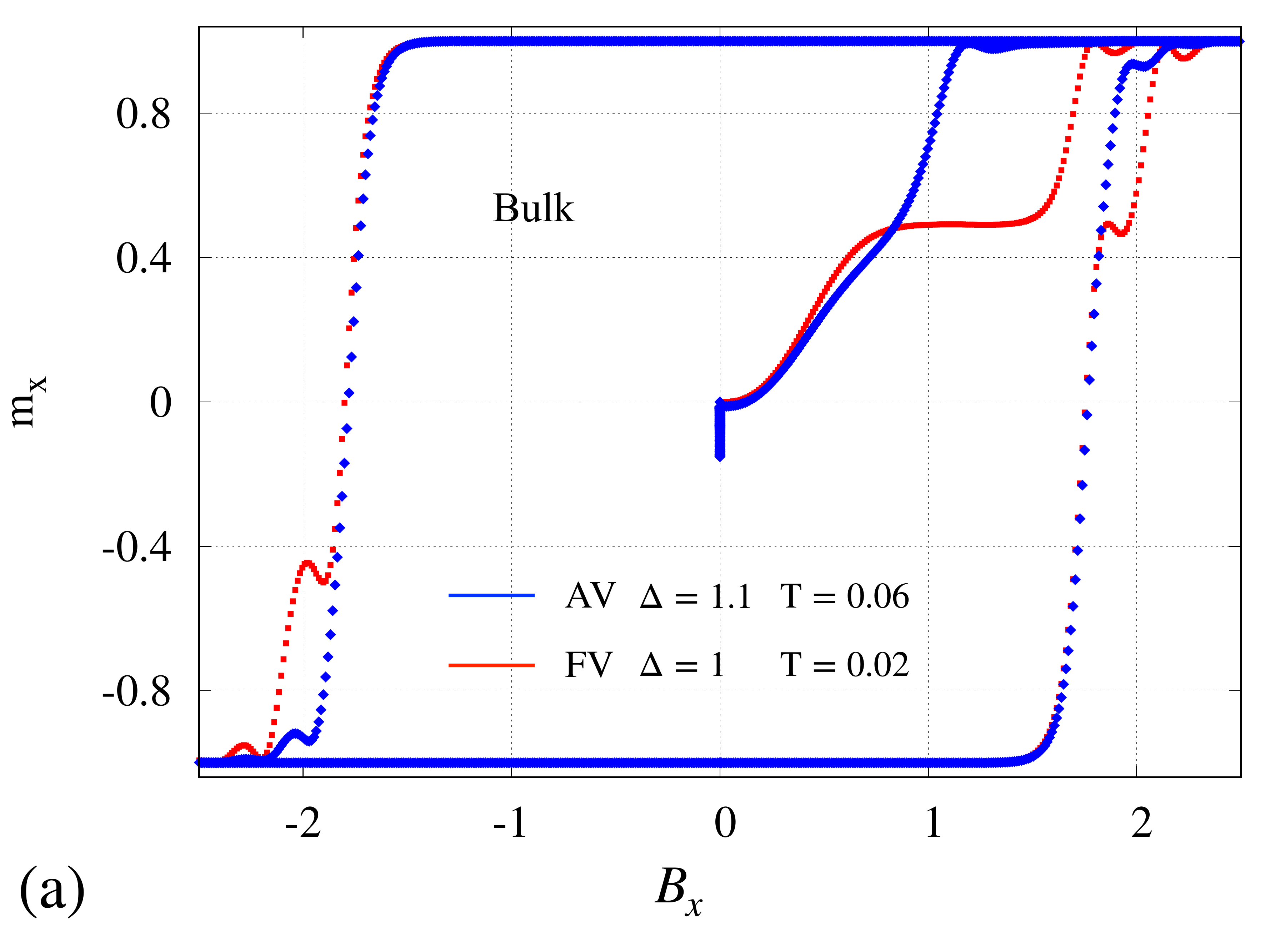}
\includegraphics[width=\columnwidth]{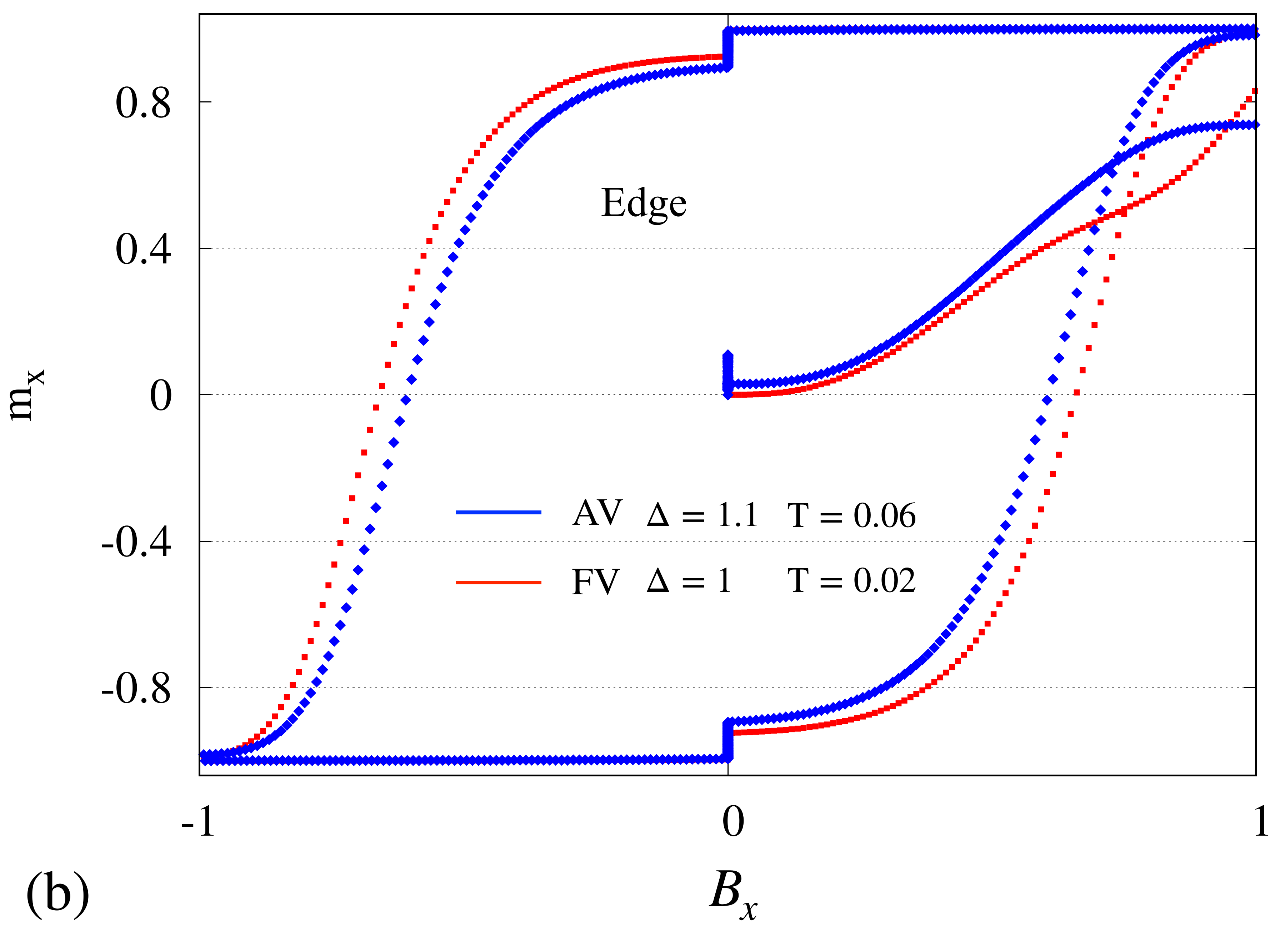}
\caption{(color online) (a) Magnetization dynamics along the direction parallel to the applied field, of dipoles in the bulk of the clusters in the $\rm AV$ (blue) and $\rm FV$ (red) magnetic orders. (b) likewise (a) but for dipoles located at the edge. In all cases $\rm m_x$ is in $\rm m_0$ units and the magnetic field $B_x$ is in $\frac{\mathcal{K}(\Delta)}{m_0}$ units.}
\label{fig:f5}
\end{figure}
\subsection{Edges versus bulk}
\label{sec:Edges versus bulk}
Aimed to compare the relaxation dynamics  of magnets located at the edge and bulk of a lattice we study a dipolar cluster made out of four square plaquettes of dipoles as shown in Fig.~\ref{fig:f3}.  From a disordered magnetic configuration, this cluster relaxes into either the antiferromagnetic vortex state denoted $\rm AV$ and shown in Fig.~\ref{fig:f3}(a) or the ferromagnetic vortex state denoted $\rm FV$ and shown Fig.~\ref{fig:f3}(b) (along with their respective time reversal versions). Since in both cases the total magnetization cancels out, we use the chirality $\chi=\frac{1}{8}(\sum_k\bm{\hat{m}}_k\times\bm{\hat{m}}_{k+1})\cdot(0,0,1)$ defined as the $\bm{z}$ projection of the average vector product of two adjacent magnetic moments as a suitable order parameter to characterize the magnetic configurations. $\rm AV$ has $\chi=-\frac{1}{2}$, while for $\rm FV$, $\chi=1$. 

Whether after relaxation the system settles into $\rm AV$ or $\rm FV$ depends on $\Delta$ and T. Energetics dictates that at T=0 $\rm FV$ is slightly favored over $\rm AV$ but this difference becomes smaller as $\Delta$ grows (Appendix \ref{app:Energies} Fig.~\ref{fig:f3A}(a)). This is also apparent in Fig.~\ref{fig:f3}(c) where the total dipolar energy density of the cluster is plot as a function of $\chi$ for several values of $\Delta$.  We see that for all values of $\Delta$ the energy is minimized for magnetic states with $\chi=-1/2$ and $\chi=1$. The small energy difference between the two diminishes dramatically with $\Delta$ because the magnitude of the dipolar interactions decrease.  Finite temperatures can overcome the small energy barrier between any of the two states, because, alike in the previous case it approaches zero as $\Delta$ grows (Fig.~\ref{fig:f3A}(b)). \\
Next, we use this square cluster as a prototype model for studying the thermal dynamics of edge and bulk dipoles in dipolar arrays.  \\
Because of a lower symmetry, edge dipoles will sustain an anisotropic internal magnetic field becoming more susceptible to external perturbations than those at the bulk. To illustrate this point, consider the two magnets highlighted in red, at the edge and bulk of the clusters shown in Figs.~\ref{fig:f3}(a),(b). The (red) magnet at the bulk of Fig.~\ref{fig:f3}(a) senses a net magnetic field parallel to its magnetic moment from its nearest collinear dipole (located at its left), because the field due to all other magnets cancels out. The (red) edge dipole at the left bottom corner of the lattice, sustains the field due to its nearest collinear magnet but this is attenuated by the field contributions from the other four parallel dipoles at the rows above. Therefore an edge dipole sustains a lower internal field and it is more unstable respect to external perturbations that a dipole at the bulk. The scenario is such that when the system is subject to external fields, dipoles at the edge of the lattice respond faster to the external torque than dipoles at the bulk.  A similar situation occurs for the dipoles highlighted in red in Fig.~\ref{fig:f3}(b).  \\
Indeed, the anisotropy of the internal magnetic field in a lattice is captured by the geometrical factor $\Lambda$ which in the square cluster splits into $\Lambda^{(\rm E)}=(6\sqrt{2}-1)$ and $\Lambda^{(\rm B)}=\frac{\Lambda^{(\rm E)}+1}{2}$ splitting $\mathcal{K}$ into $\mathcal{K}^{(\rm E)}$ and $\mathcal{K}^{(\rm B)}$ for edge and bulk states respectively. The splitting of $\mathcal{K}$ affects the relaxation dynamics through $\epsilon$, $\tau_2$ and $\tau_3$. The final result is the shifting of $\mathcal{C}(s)$ for magnets at the edge and the bulk.  This scenario is verified in Fig.~\ref{fig:f3}(d) which shows the evolution of $\mathcal{C}(s)$ of bulk (curve in blue) and edge (curve in red) dipoles after applying a small perturbation to a cluster that originally relaxed into the $\rm FV$ state (with $\Delta=1.1$ and at $\rm T=0.06$). As expected,  numerical simulations  (main figure) show that the dynamics of dipoles at the edge and bulk is shifted and that edge states evolve faster than bulk magnets. Furthermore the numerical solution captures a qualitative difference between the thermal relaxation in both cases: while edge dipoles evolve in a fashion reminiscent of  Fig.~\ref{fig:f2}(a), bulk magnets evolve in a smother manner. The inset corresponds to the evaluation of Eq.~(\ref{eq:auto}) using $\mathcal{K}^{(\rm E)}$ and $\mathcal{K}^{(\rm B)}$ for edge and bulk dipoles respectively which, for short times, yields the same qualitative behavior that the main figure. 
 \subsection{Magnetization dynamics}
 \label{sec:Magnetization dynamics}
Finally we investigate the thermal dynamics of the square dipolar clusters under a uniform external magnetic field. To that effect molecular dynamics simulations are used to solve the equation:
\begin{eqnarray}
I\frac{d^2\alpha^i}{dt^2}=\sqrt{2\eta k_B T}\xi(t)-\mathcal{\eta}\frac{d\alpha^i}{dt}-\mathcal{T}^i_z -\mathcal{T}^{(\rm{i,e})}_z
\label{eq:langevinb}
\end{eqnarray}
where $\mathcal{T}^{(\rm{i,e})}_z=\bm{m}^i\times\bm{B}$ and $\bm{B}$ denotes a uniform magnetic field applied in the $x-y$ plane. 

We prepared two systems by solving the thermal relaxation (Eq.~(\ref{eq:langevin1})) of one cluster  with $\Delta=1$ and at $\rm T=0.02$ and a second one with $\Delta=1.1$ and at $\rm T=0.06$. In the first case the system relaxed in the $\rm FV$ and in the second it settled into the $\rm AV$ state. Next Eq.~(\ref{eq:langevinb}) was numerically solved for each of them. The resulting  magnetization along the $x$ direction, $\rm m_x$ (in units of $\rm m_0$), due to an external magnetic field applied along the $x$ axis, $\rm B_x$ (in units of $\rm \mathcal{K}(\Delta)/m_0$) is shown in the respective blue and red curves of Fig.~\ref{fig:f4}. We note that the magnetization loops depict similar behavior in both lattices. The plateaux at $m_x=\frac{1}{3}$ realized in both cases deserves special attention. To inspect it further, the loop of Fig.~\ref{fig:f4} is broken up into  the magnetization of dipoles at the bulk and at the edge of the clusters as shown in Figs.~\ref{fig:f5}(a) and (b) respectively.  Now the dissimilar magnetization dynamics of bulk and edge is apparent in both clusters (the $\rm AV$ is shown in blue and the $\rm FV$ is shown in red). Consequently, the $\frac{1}{3}$ feature is attributed  to the anisotropy of the internal dipolar interactions between bulk and edges dipoles as discussed in Subsection \ref{sec:Edges versus bulk}.  While dipoles at the edge, Fig.~\ref{fig:f5}(b), respond easily to very small values of $B_x$  the four dipoles at the bulk of the lattices Fig.~\ref{fig:f5}(a) stay pinned until at $B_x\sim1.8$ they suddenly rotate to follow the direction of the external field. Since dipoles in the bulk correspond to one fourth of the total number of magnets in the system, their action leaves a signature in the form of a plateaux in the magnetization loop. Therefore, the width of the plateaux of Fig.~\ref{fig:f4} is a measure of the anisotropy between the internal fields at the edge and at the bulk of the sample. The small shoulder at $B_x\sim2$ in the red curves  of Figs.~\ref{fig:f4},~\ref{fig:f5} is due to the slightly delayed flip of one of the bulk dipoles as shown in the supplementary videos  \cite{supp}.
\section{Conclusions}
\label{sec:Conclusions}
Square clusters of magnetic dipoles have been studied as prototype  models to elucidate the role of internal dipolar interactions, intrinsic  properties of the magnetic degrees of freedom and the geometric  features of a lattice, in the thermal relaxation and magnetization dynamics of dipolar arrays. By solving the Langevin equation for the angular rotation of interacting dipoles we found that the early relaxation dynamics of the systems under study is determined by temperature, dipolar interactions and inertia while the long time relaxation is defined by the interplay between damping and magnetic couplings. Temperature, magnitude of dipolar interactions, damping coefficient and inertial aspects of the magnets, are imprinted in the time scales that determine the stages of evolution of the time autocorrelation function of an array of magnets. 
 The study of the Langevin dynamics allows to set apart geometrical aspects of the lattice from the magnetic and inertial properties of the dipoles.  Consequently we define a magnetic correlation length $\ell$ in terms of  inertia, damping and magnetic intensity of the spins while the symmetry aspect of the array is stored in a geometric factor which is lattice dependent. For the case of nanoarrays of  mesospins,  $\ell$ could be a useful length scale to compare with the lattice constant aimed to determine whether or not internal correlations play a dominant role in the dynamics of the system at hand. 
 
The anisotropy of the internal magnetic fields in a lattice is captured by a geometrical factor $\Lambda$ which distinguishes the magnetic torques sustained by dipoles at the bulk and at the edge of a lattice. The magnetic anisotropy of the internal fields manifests in the proper time scales of the system which differ for dipoles at the bulk and the edge. This has consequences in the time autocorrelation function which shows qualitative differences for edge and bulk magnets. When such a system is under an external magnetic field, signatures of the dipolar anisotropy are displayed in the magnetization dynamics though a plateau that shows up in the magnetization loops of the clusters. Sorting out the magnetization reversal of edge and bulk magnets, reveals once again the qualitative differences in the magnetization dynamics of edge and bulk states in dipolar arrays.  
\section*{Acknowledgments}
This work was supported in part by Fondecyt under Grant No. 11121397.  The author acknowledges support from the Simons Foundation and thanks Professor Vassilios Kapaklis for sharing experimental data that motivated this work.
\appendix
\section{Solution of Eq.~(\ref{eq:langevin2})}
\label{app:Solution of Langevin equation}
The Langevin equation that determines the dynamics of the angular variable $\alpha^i$ is \cite{ullersma1966exactly,bojdecki1991gaussian,pathria2011statistical}: 
\begin{eqnarray}
I\frac{d^2\alpha^i}{dt^2}=\sqrt{2\eta k_B T}\xi(t)-\mathcal{\eta}\frac{d\alpha^i}{dt}-\mathcal{K}\alpha^i,
\label{eq:langevin}
\end{eqnarray}
where $ I$ ([Kg $\rm m^2$]) is the inertia moment of each dipole, and $\eta$ ([$\frac{\rm Kg  m^2}{s}$]) is a damping coefficient that accounts for its viscous rotation in the $x-y$ plane. Thermal fluctuations due to the coupling of the magnet with the thermal bath are modeled by a $\delta$-correlated Gaussian noise $\xi(t)$ of zero mean and unit intensity: $\langle \xi(t)\rangle=0,$ $\langle \xi(t) \xi(t')\rangle=\delta(t-t')$.  T denotes temperature and $k_B$ is the Boltzmann constant. 
The  last term $\mathcal{K}\alpha^i$ accounts for the torque on dipole $\bm{m}_i$ due to the internal magnetic field originated by all other dipoles assuming that 1) dipoles deviate slightly from their equilibrium positions and 2) at a given position the internal fields perpendicular and parallel to $\bm{m}_i$ are such that $B_{\bot}\ll B_{||}$ , yields $\mathcal{K}=m_0 B_{||}$ as explained in Subsection \ref{sec:Interacting case}.\\
Hereafter the superscript $i$ will be omitted. 
Consider
\begin{equation}
\alpha(t)=\alpha_h(t)+\alpha_\xi(t), 
\end{equation}
with $\alpha_h(t)$ solution of the homogeneous equation $I\ddot{\alpha}_h=-\eta\dot{\alpha}_h-\mathcal{K}\alpha_h$. $\alpha_h(t)$ is a linear combination of two independent solutions: $\alpha_h=A \rm{e}^{\eta_0 t}\sin{\omega t}+B\rm{e}^{\eta_0 t}\cos{\omega t}$, where $A$ and $B$ are constants to be determined from $\alpha_h(0)$ and $\dot{\alpha_h}(0)$. In addition  $\omega^2=\omega_0^2-\eta_0^2$, $\omega_0^2=\frac{\mathcal{K}}{I}$ and $\rm \eta_0=\frac{\eta}{2I}$.

On the other side, $\alpha_\xi$  is a particular solution related to  $\xi(t)$ and satisfies the inhomogeneous equation with initial conditions $\alpha_\xi(0)=\alpha_0$ and $\dot{\alpha}_\xi(0)=v_0$. 

$\alpha_\xi$ can be expressed by the Green function,
\begin{equation}
\alpha_\xi(t)=\sqrt{2\eta k_B T}\int_0^\infty\mathcal{G}(t,u)\xi(u)du
\end{equation}
For obtaining $\mathcal{G}(t,u)$ we use the solution of the following homogeneous equation:
\begin{equation}
I\ddot{\alpha}+\eta\dot{\alpha}+\mathcal{K}\alpha=0
\label{eq:homo}
\end{equation}
the solution of Eq.~(\ref{eq:homo}) takes the form: $\alpha(t)=e^{rt}$, which once replaced in the homogeneous equation yields:
\begin{equation}
 I r^2+\eta r+\mathcal{K}=0
\end{equation}
Giving two real solutions for r: $r_{\pm}=-\frac{\eta}{2I}\pm\beta$ with $\beta=\frac{\eta}{2I}\sqrt{1-\frac{4\mathcal{K} I}{\eta^2}}$. Therefore $\alpha_1(t)=\exp\left[(\beta-\frac{\eta}{2I})t\right]$ and $\alpha_2(t)=\exp\left[-(\beta+\frac{\eta}{2I})t\right]$.\\
Next we construct the Green function that verifies:
\begin{equation}
I\ddot{\mathcal{G}}(t,t') +\eta\dot{\mathcal{G}}(t,t') +\mathcal{K}\mathcal{G}(t,t') =\delta(t-t^{'})
\label{eq:langevin4}
\end{equation}
with the initial conditions $\mathcal{G}(0,t')=0$ and $\dot{\mathcal{G}}(0,t')=0$. $\mathcal{G}$ can be written as a linear combination of solutions of the homogeneous equation as follows:
\begin{eqnarray}
\mathcal{G}(t,t') =c_1\alpha_1+c_2\alpha_2, \hspace {12pt} t<t' \\ \nonumber
\mathcal{G}(t,t') =d_1\alpha_1+d_2\alpha_2, \hspace {12pt} t>t'
\end{eqnarray}
The constants $c_1$, $c_2$, $d_1$ and $d_2$ can be determined from the initial conditions and the continuity of $\mathcal{G}(t,t')$ at $t=t'$. They imply that $c_1=c_2=0$ and $c_1\alpha_1(t')+c_2\alpha_2(t')=d_1\alpha_1(t')+d_2\alpha_2(t')$ which yields $d_1\alpha_1(t')=-d_2\alpha_2(t')$.\\
Integrating  Eq.\ref{eq:langevin4} from $t^{'+}$ to $t^{'-}$ we obtain,
\begin{equation}
\int_{t^{'-}}^{t^{'+}}\left[I\ddot{\mathcal{G}}(t,t') +\eta\dot{\mathcal{G}}(t,t') +\mathcal{K}\mathcal{G}(t,t') \right]dt=\int_{t^{'-}}^{t^{'+}}\delta(t-t^{'})dt
\end{equation}
Since $\mathcal{G}(t,t')$ is continuous, $\dot{\mathcal{G}}(t,t')$ can have only a jump discontinuity and therefore
\begin{equation}
\dot{\mathcal{G}}(t,t')|_{t=t^{'+}}-\dot{\mathcal{G}}(t,t')|_{t=t^{'-}} =\frac{1}{I}
\end{equation}
Yielding,
\begin{eqnarray}
d_1(t') =\frac{1}{2I\beta\alpha_1(t')}\\ \nonumber
d_2(t') =\frac{-1}{2I\beta\alpha_2(t')}
\end{eqnarray}
and the Green function becomes
\begin{equation}
\mathcal{G}(t,t')=\Theta(t-t')\frac{1}{2 I \beta}\rm e^{-\frac{\eta}{2I}(t-t')}\left[e^{\beta(t-t')}-e^{-\beta(t-t')}\right]
\end{equation}
Since $\xi(t)$ is a gaussian process, $\alpha_\xi(t)$ is gaussian too and then $\alpha(t)$ is a gaussian stochastic process. Because $\langle \xi(t)\rangle=0$, it follows that $\langle\alpha_\xi(t)\rangle=0$. Because $\alpha(t)=\alpha_h(t)+\alpha_\xi(t)$, the mean square deviation of the particle angle 
\begin{align}
\langle(\delta\alpha(t))^2\rangle=\langle\alpha^2(t)\rangle-\langle\alpha(t)\rangle^2
\end{align}
using the Green function can now be written as:
\begin{align}
(\alpha(t))^2&=&\frac{2\eta k_B T}{I^2\omega^2}\int_0^t\int_0^t dt'dt^{''}e^{-\eta_0(2t-t'-t^{''})}\nonumber \\&&\sin{\omega(t-t')}\sin{\omega(t'-t^{''})}\xi(t')\xi(t^{''})
\end{align}
Taking average over noise realizations yields \cite{yaghoubi2017energetics}:
\begin{align}
\langle(\delta\alpha(t))^2\rangle&=&\langle(\delta\alpha_\xi(t))^2\rangle=\frac{2\eta k_B T}{I^2\omega^2}\int_0^tdt'e^{-2\eta_0(t-t')}\nonumber \\&&\sin^2{\omega(t-t')}
\label{eq:angle}
\end{align}
Finally evaluation of the integral in Eq.(\ref{eq:angle}) leads to the mean square deviation of the angle of a dipole:
 \begin{align}
\langle(\delta\alpha(t))^2\rangle=\frac{k_B T}{\mathcal{K}}+\frac{k_B T}{\mathcal{K}\omega^2}\left[\eta_0^2\cos{2\omega t}-\eta_0\omega\sin{2\omega t}-\omega_0^2\right]
\end{align}
\section{Molecular dynamics simulations of the Langevin dynamics}
\label{app:Molecular dynamics simulations of the Langevin dynamics}
Numerical results were obtained by direct numerical integration of the Langevin equation of motion for each magnet interacting with all the others via dipolar interactions, Eq.(\ref{eq:langevin1}) or Eq.(\ref{eq:langevinb}) depending on the case.  
To solve the system of equations, we use a Verlet method with an integration time step $d t=2\times10^{-6}$ [s]  equivalent to $\sim 2.5\cdot10^{6}$ time steps. In all numerical simulations, the time discretization step satisfied the condition $dt<10^{-2}\text{Min}\{ \tau_{1}, \tau_{2}, \tau_{3},\tau_{th}  \}$. 

For  the square plaquette and for the cluster, energy minimization and molecular dynamics simulations run at T=0 from an initial random magnetic configuration,  yielded the minimum energy magnetic state. Next the system was initialized such that  each dipole was slightly taken away from its equilibrium orientation by a small random amount.  Then using the Verlet algorithm the system of dipoles follows the dynamics modeled by Eq.(\ref{eq:langevin1}) and is left to reach equilibrium for $ \sim 5$ seconds.

\begin{figure}
 \includegraphics[width=\columnwidth]{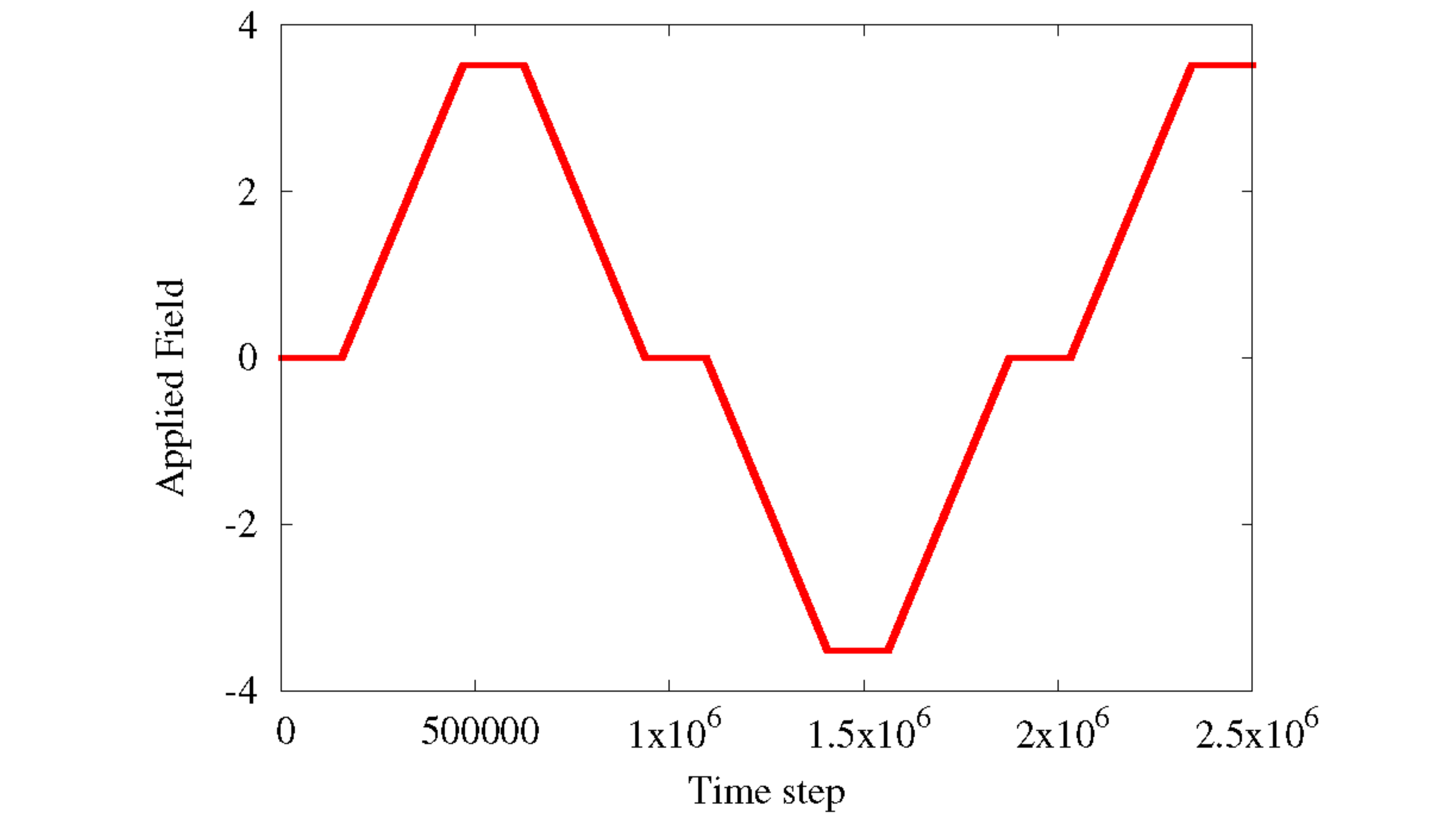}
\caption{Magnetic field ramp used in numerical simulations in units of $\mathcal{K}(\Delta)/m_0$.}
\label{fig:f1A}
\end{figure}

For the magnetization dynamics analysis under an external field, we used a uniform magnetic field along the $\bm x$ direction (Eq.(\ref{eq:langevinb})), which changed by $\delta \rm{B} \sim 10^{-5}$ in each time step. During the simulation interval, the magnitude of the field increased from  0 up to $\rm{B}=\rm{B^{max}}$, next it went back to zero to decrease down to $-\rm{B^{max}}$. Then it returned to zero to finally rise up to $\rm{B^{max}}=3.51$ as shown in Fig.\ref{fig:f1A}. The total simulation time was $5$ s. In all data presented here magnetic fields are normalized by $\mathcal{K}(\Delta)/m_0$.  In all simulations, the length of the magnets is $L=1$, the moment of inertia $I=\mathcal{K}(\Delta)\times10^{-3}$, the damping $\eta=50 I$, saturation magnetization $M_s=10^6/\pi$ [$\rm A/m$], the radius of the magnets  $r=L\times10^{-3}$, the magnetic charge $\rm q=M_s\pi r^2=1$, $m_0=L q=1$ and finally the characteristic length scale $\ell=(\frac{\rm \mu_0 m_0^2  I}{\eta^2})^{1/3}=1.04$. To produce $\xi(t)$, a given temperature T was multiplied by a random number with a gaussian distribution produced by Mathematica 12.0. routine $\it{RandomVariate[NormalDistribution[]]}$ \cite{Mathematica}. The range of temperatures spanned was $(10^{-2},1)$.
\begin{figure}[h]
 \includegraphics[width=\columnwidth]{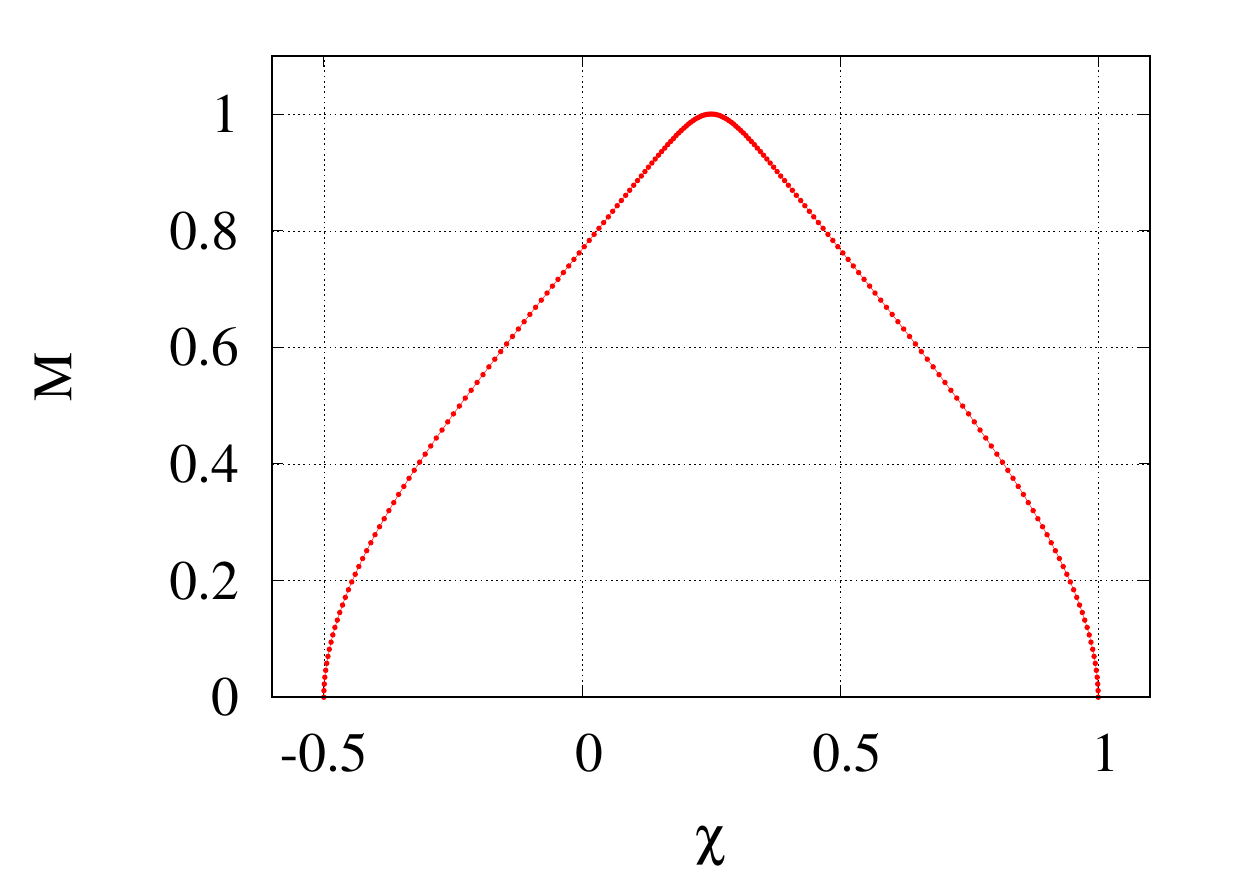}
\caption{Normalized magnetization of the cluster versus $\chi$ at $T=0$. }
\label{fig:f2A}
\end{figure}
\begin{figure}
 \includegraphics[width=\columnwidth]{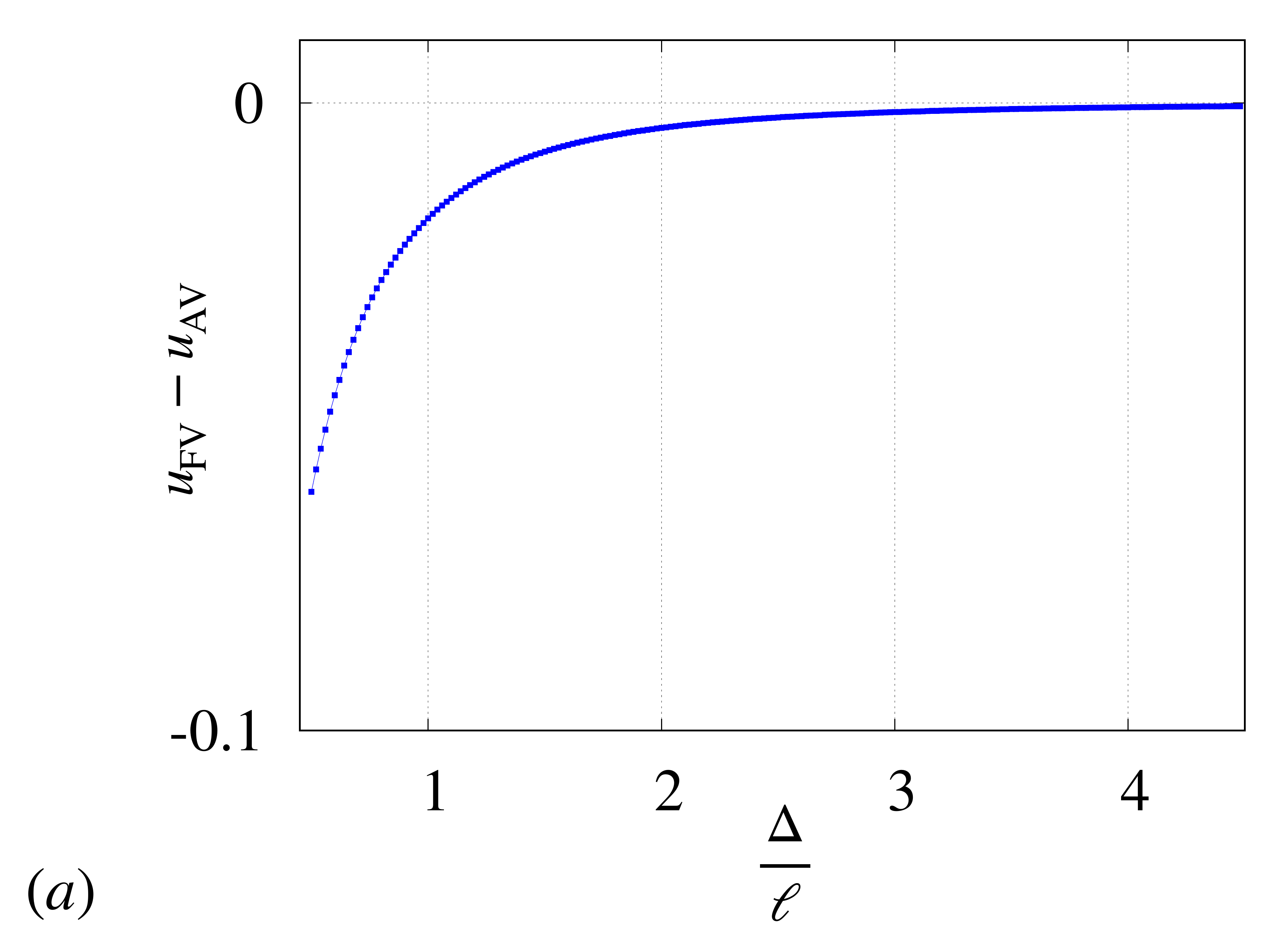}
 \includegraphics[width=\columnwidth]{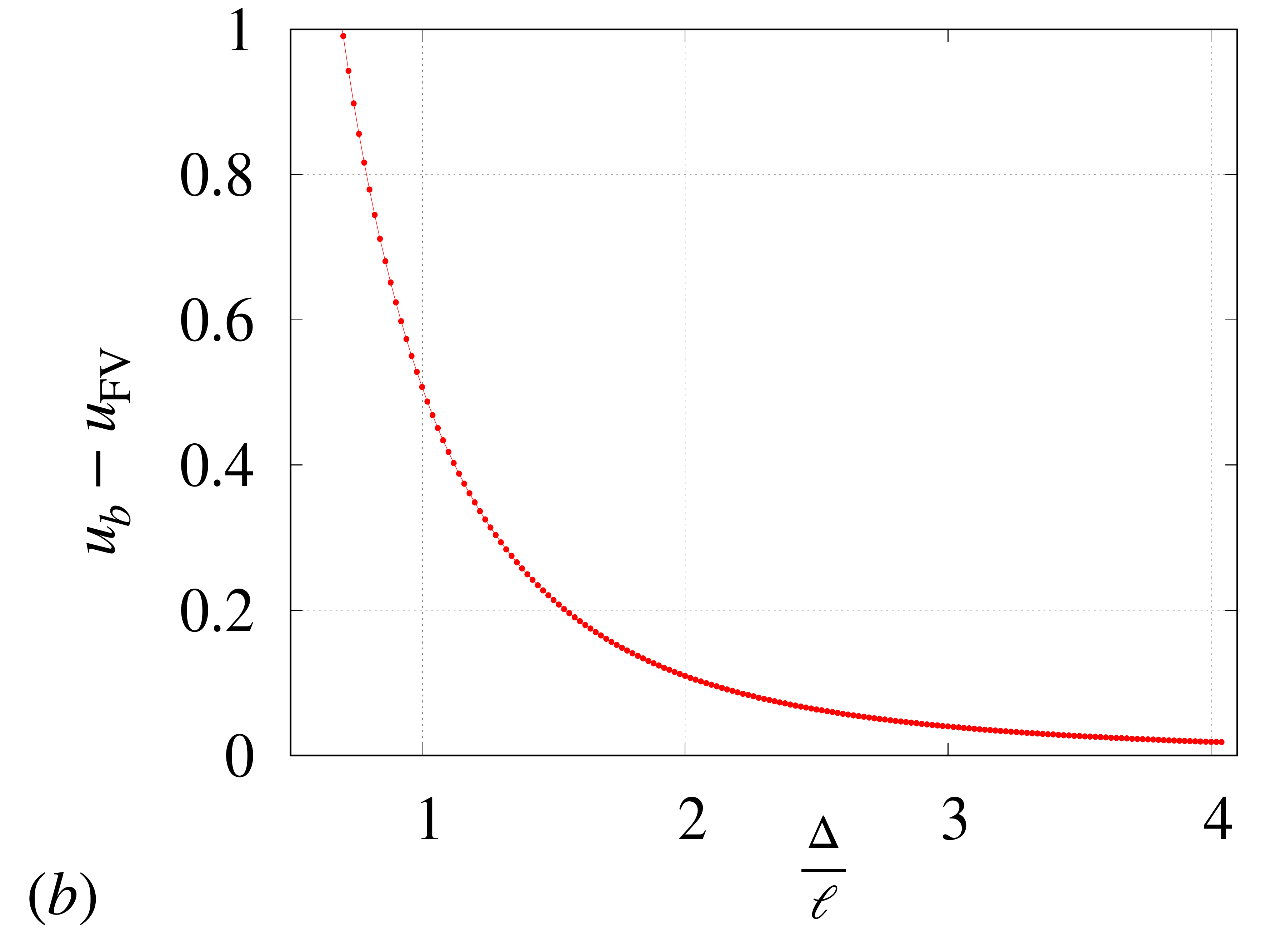}
\caption{(a) Difference between the dipolar energy of the magnetic configurations $\rm FV$  and  $\rm AV$ and (b) Difference between the maximum energy density of the system (the energy barrier reached when $M=1$ and $\chi=0.26$) and the energy of the magnetic configuration $\rm FV$ in terms  of $\Delta$ at $T=0$. All energies are normalized by $\mathcal{K}(L)$.}
\label{fig:f3A}
\end{figure}
 \begin{figure}
 \includegraphics[width=\columnwidth]{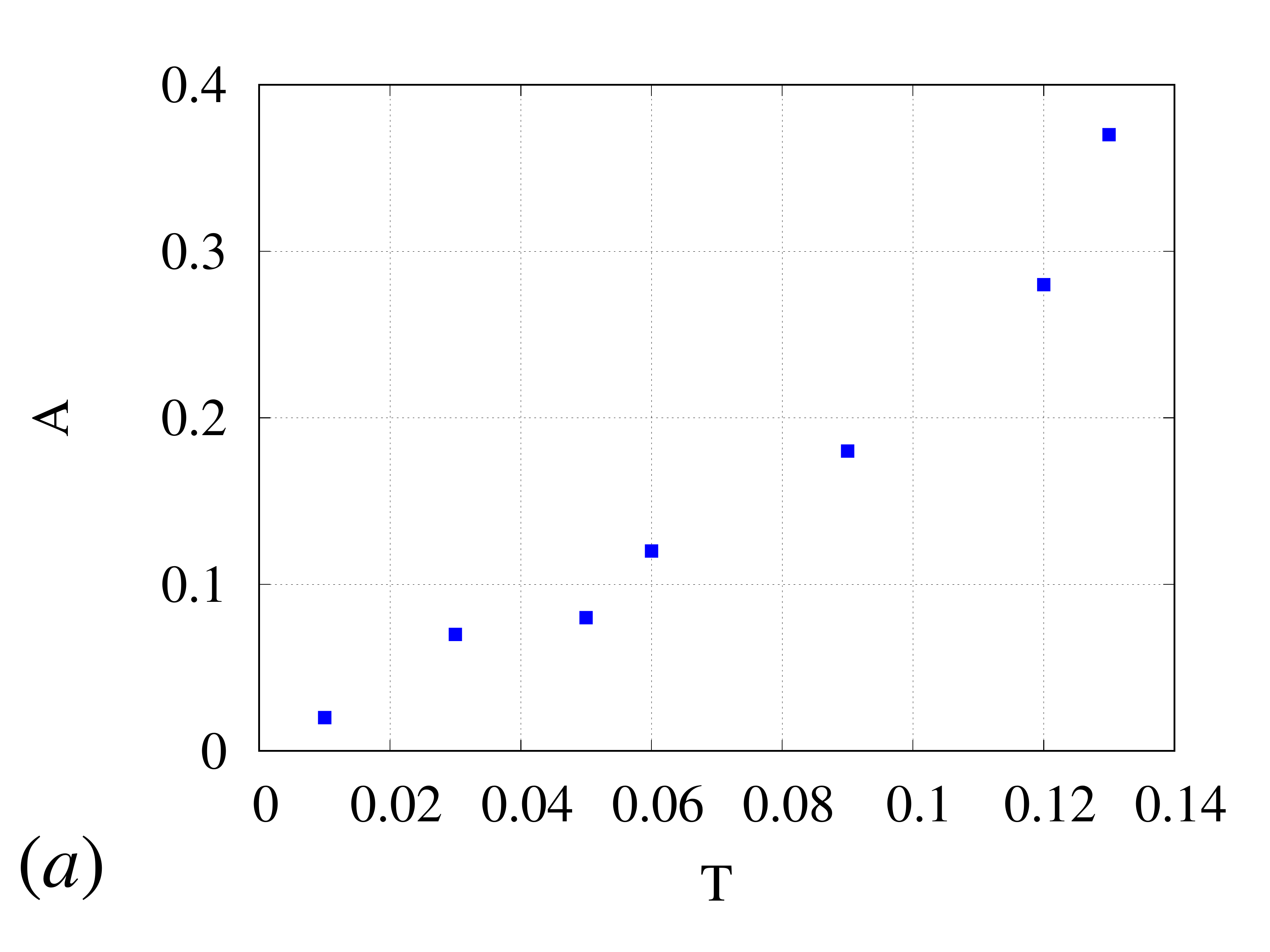}
 \includegraphics[width=\columnwidth]{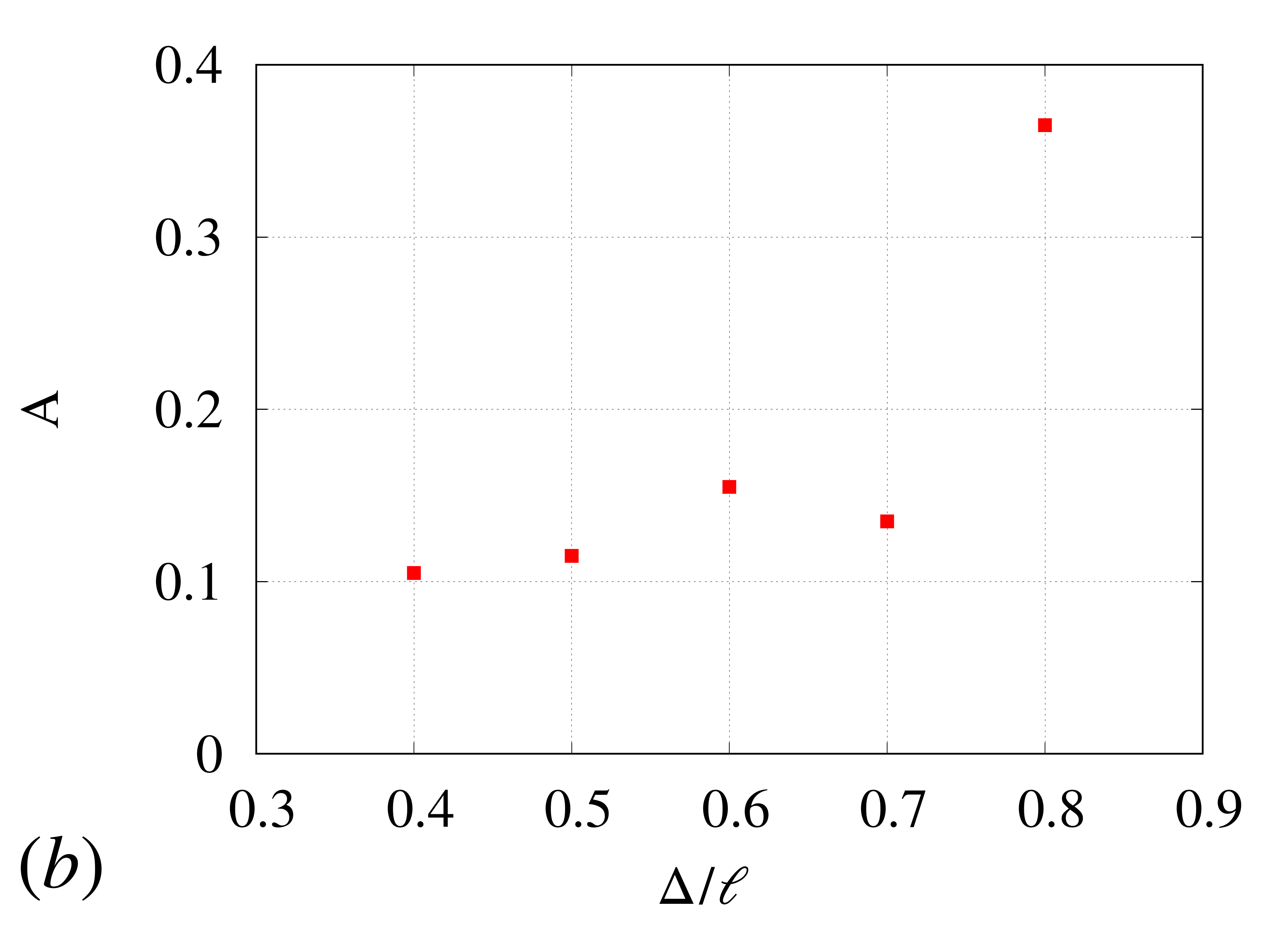}
\caption{(a) The amplitude $A$ of the earliest oscillation of $\mathcal{C}(s)$ of a single square plaquette of dipoles is shown as a function of temperature T. (b) $A$ in terms of $\Delta$. (a) and (b) are obtained from the numerical solution of Eq.(\ref{eq:langevin1}). T is normalized by $\mathcal{K}(L)$.}
\label{fig:f4A}
\end{figure}
\section{Energies of $\rm AV$ and $\rm FV$ at T=0}
\label{app:Energies}
The equilibrium configuration of the cluster after relaxation from a random magnetic configuration depends on $\Delta$ and T.  Energetics dictates that at $T=0$ the $\rm FV$ is slightly favorable for all values of $\Delta$, but decreases as $\Delta$  grows as shown in Fig.\ref{fig:f3A}(a). Here we show the difference between the dipolar energy of the cluster divided by the number of dipoles and normalized by $\mathcal{K}(L)$ versus $\Delta/\ell$. We see that the difference between the energy of the two states converges in a logarithmic fashion with the growing of $\Delta$. 

A similar situation occurs when we examine the difference between the energy of $\rm FV$ and the energy barrier between the two equilibrium configurations. The energetic barrier is set by the maximum dipolar energy of the cluster which occurs when it realizes a magnetic configuration that has maximum magnetization $M=\frac{1}{12}\sum_k|\bm{\hat{m}}_k|=1$ and has $\chi=0.26$ as shown in Fig.\ref{fig:f3A}(b)  and Fig.\ref{fig:f3}(c) respectively. We can see that the effect of increasing $\Delta$ is to decrease the dipolar energy of the system which consequently decreases the energy barrier and the energy difference between the equilibrium configurations in the cluster.
\section{$A$ versus T and $\Delta$}
\label{app:amplitude}
In Figs.\ref{fig:f4A}(a) and \ref{fig:f4A}(b)  we show the amplitude $A$ of $\mathcal{C}(s)$ at the onset of the relaxation dynamics in terms of T and $\Delta$ respectively. As expected Fig.\ref{fig:f4A}(a) shows the increment of $A$ with temperature. 
As $\Delta$ grows the strength of the dipolar interactions between magnets decreases which means that inertial effects take over. This is realized by the augment of A as $\Delta$ grows as shown in Figs.\ref{fig:f4A}(b).    

\end{document}